\documentstyle[aps,epsfig,amssymb,amsmath,manuscript]{revtex}

\title{Fluctuation superconductivity limited noise in a transition-edge sensor}

\author{A. Luukanen, K. M. Kinnunen, A. K. Nuottaj\"arvi \\ 
Department of Physics, University of Jyv\"askyl\"a \\ P.O.Box 35 (YFL) FIN-40014 University of Jyv\"askyl\"a, Finland \\
H. F. C. Hoevers, W. M. Bergmann Tiest \\
 SRON National Institute for Space Research\\ Sorbonnelaan 2, 3584 CA Utrecht, The Netherlands\\
J. P. Pekola \\ 
Low Temperature Laboratory, Helsinki University of Technology \\ P.O. Box 2200, FIN-02015 HUT, Finland 
}

\begin{document}

\maketitle

\begin{abstract} 
{In order to investigate the origin of the until now unaccounted excess noise and to minimize the uncontrollable phenomena at the 
transition in X-ray microcalorimeters we have developed superconducting transition-edge sensors into an edgeless geometry, the 
so-called Corbino disk (CorTES), with superconducting contacts in the centre and at the outer perimeter. The measured rms current 
noise and its spectral density can be modeled as resistance noise resulting from fluctuations near the equilibrium 
superconductor-normal metal boundary.} 
\end{abstract}

At present, the most sensitive energy-dispersive X-ray detector is the transition-edge sensor (TES) microcalorimeter, a thermal 
detector operated typically at a bath temperature below 100 mK \cite{irwin1,wouter1,luukanen4}. The device  consists of an X-ray 
absorber (Bi, Au, Cu being the most common materials), thermally coupled to a TES superconducting film with a critical temperature 
$T_{\rm c}\approx 100$ mK. The TES film is typically a proximity-coupled bilayer, e.g., Ti/Au, Mo/Cu, or Mo/Au. A common detector 
geometry is a square TES film, covered completely, or in some cases partially by the X-ray absorber film. The TES film - absorber 
combination is located on a thermally isolating Si$_3$N$_4$ film, micromachined to a bulk Si substrate which acts as the heat sink. 
Wires with a $T_{\rm c}$ much higher than that of the TES film are used to connect the detector to the bias circuit. The device is 
connected to a constant voltage bias, and the current through the sensor is measured with a superconducting quantum interference 
device (SQUID). The theory of operation of these devices has been well developed, but is not complete as the TES microcalorimeters 
consistently do not achieve the energy resolution predicted by the models. Firstly, the TES microcalorimeters fail to reach the 
expected energy resolution in calorimetry, especially when the deposited heat drives the device through a large part of its 
superconducting transition. Secondly, most TESs exhibit noise in excess of the sum of the commonly recognized noise components: 
thermal fluctuation noise arising from the thermal link between the TES and the heat sink (TFN), Johnson noise (JN), and SQUID 
(read-out) noise (SN). This letter presents a simple model which explains this discrepancy in the detector subject to our study.


In the square devices, edges parallel to the current become crucial for the device performance. Firstly thickness variations 
resulting from underetching or imperfect deposition of the TES bilayer lead to spatial $T_{\rm c}$ variations. Secondly the edges 
have also proven to give rise to flux creep noise with the higher concentration of trapping centres due to local defects which can 
be observable at certain values of the bias voltage. A solution to overcome edge effects is to deposit thick normal metal banks over 
the edges. The proximity effect of the thick normal metal reduces the critical temperature of the edges well below that of bulk of 
the TES film, resulting in well-defined edges \cite{hilton1}.

\begin{figure} 
\begin{center}
\includegraphics[width=12cm,angle=0]{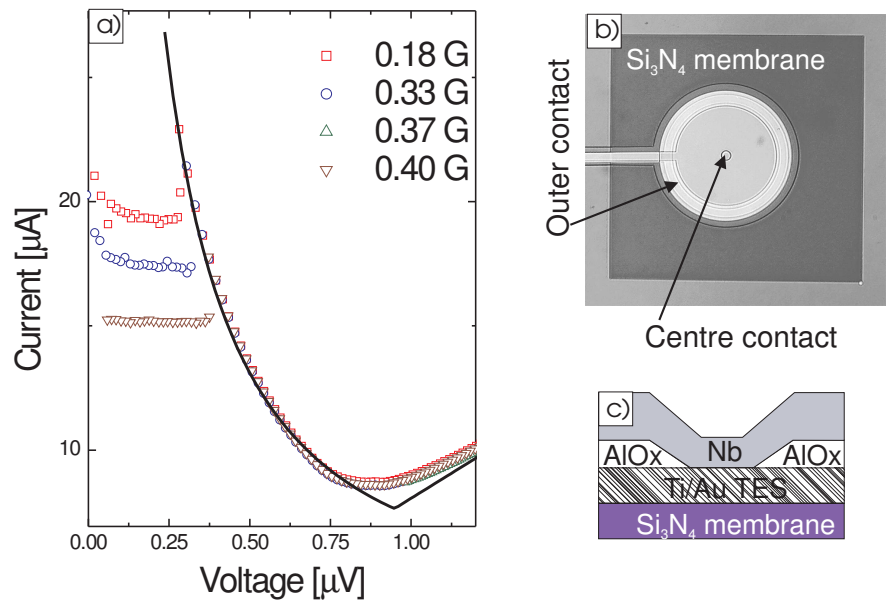}
\vspace{2cm}
\caption{A. Luukanen, H.F.C. Hoevers, K.M. Kinnunen, A.K. Nuottaj\"arvi, J.P. Pekola, and W.M. Bergmann Tiest}
\label{fig:cortes_fig1}
\end{center} 
\end{figure}

Another way of removing the edges is to use a Corbino disk geometry, in which a current source is placed in the apex of the annular 
TES film, and another superconducting contact is placed to the outer circumference of the TES film. Here the current density is 
proportional to $1/r$, which results in a well defined phase boundary at certain distance $r_{\rm b}$ from the centre of the disk. A 
Corbino geometry TES, or the CorTES, is shown in the inset of Fig. \ref{fig:cortes_fig1}. The central contact is provided by a 
superconducting ground plane, covering the entire sensor. By this we ensure a truly cylindrical symmetry and a homogeneous current 
distribution with a concentric current return. 

The devices are fabricated on a double-side nitridized, 525 $\mu $m thick Si wafer. Free standing Si$_3$N$_4$ membranes with a 
thickness of 250 nm are fabricated by wet etching of the Si. The CorTES layers are patterned by e-beam lithography, combined with 
UHV e-beam evaporation and lift-off. The wiring layers consist of a circular Nb outer contact, and a Nb ground plane, which contacts 
the TES film through an opening in an underlying insulator.

In contrast to a square TES, the phase boundary in the CorTES evolves controllably from the centre of the disk and moves radially 
outwards with increasing current. This can be modelled by a heat transfer model, similar to that used to describe suspended Nb 
microbridge bolometers and hot electron mixers \cite{luukanen5,wilmsfloet}. Assuming radial symmetry, the current density is given 
by $j(r)=I/(2 \pi r t)$, where $I$ is the current, $r$ is the radial distance from the centre of the disk, and $t$ is the film 
thickness. Consequently, the resistance of the CorTES is given by $R=\rho_{\rm n}/(2 \pi t) \int_{r_0}^{r_{\rm b}} 1/r dr=\rho_{\rm 
n}/(2 \pi t) \ln(r_{\rm b}/r_0)$, where $r_0$ is the radius of the central superconducting contact. The steady-state behaviour can 
be modeled by first noting that the heat transport within the sensor is completely dominated by the metal films. The thermal 
conductivity in the superconducting region (at radii $r \ge r_{\rm b}$) is given approximately by $\kappa_{\rm S}(T)=\kappa_{\rm N} 
\exp[-\Lambda/k_{\mathrm B}(1/T-1/T_{\rm c})]$ where $\Lambda$ is of the order of the energy gap $\Delta$ of the superconductor, 
$k_{\rm B}$ is the Boltzmann constant, $\kappa_{\rm N}=L T_{\rm c}/\rho_{\rm n}$ is the normal state thermal conductivity, 
$\rho_{\rm n}$ is the normal state electrical resistivity and $L=2.45 \times 10^{-8}$ V$^2/$K$^2$ is the Lorentz number 
\cite{berman}. Here we assume validity of Wiedemann-Franz law. The thermal conductivity of the 250 nm thick SiN membrane is given by 
$\kappa_{\rm  M} \simeq 14.5 \times 10^{-3}T^{1.98}$ \cite{leivo2}, and at temperatures present in the system (20 mK - 150 mK) it is 
typically three orders of magnitude smaller than $\kappa_{\rm S}$. As the thicknesses of the SiN and the TES are comparable, the 
problem reduces to two dimensions, and due to symmetry further to one dimension. At this point we neglect the temperature gradient 
within the normal state part, as the gradient over the superconducting annulus and especially over the surrounding membrane are much 
larger. In the superconducting part we require a heat balance $\dot Q/(2 \pi t) \int_{r_{\rm b}}^{r_1}1/r \; dr=-\int_{T_{\rm 
c}}^{T_1} \kappa_{\rm S}(T) \;dT$, where $\dot Q=V^2 2 \pi t/\rho_{\rm n} \ln(r_{\rm b}/r_0)$ is the dissipated bias power, with $V$ 
the applied voltage across the sensor and $r_1$ the radius of the CorTES outer edge at temperature $T(r_1)=T_1$. A similar equation 
can be written for the heat transport in the membrane but now the integration is carried out from $r_1$ to $r_0=2w/\pi$, the radius 
of the "equivalent" circular membrane to the square one with a pitch of $w$, and in temperature from $T_1$ to $T_0$, the latter 
being the bath temperature. This leads to a solution for $T_1$, which can be inserted into the heat balance equation of the 
superconducting region. This can then be numerically solved for $r_{\rm b}(V)$ from which one obtains $I(V)=V/R=2 \pi t 
V/\{\rho_{\rm n} \ln[r_{\rm b}(V)/r_0]\}$.

\begin{figure} 
\begin{center}

\includegraphics[width=12cm,angle=0]{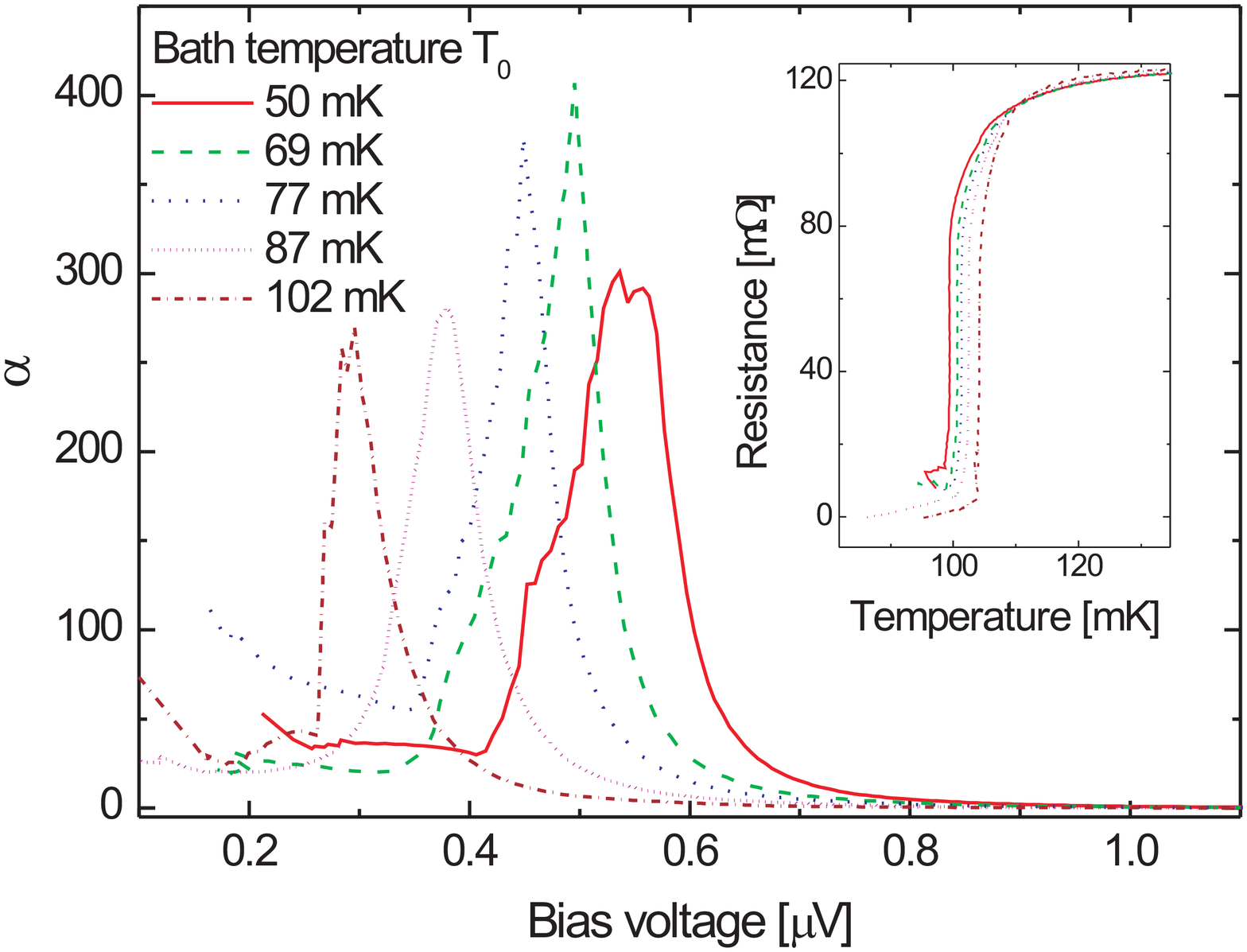}
\vspace{2cm}
\caption{A. Luukanen, H.F.C. Hoevers, K.M. Kinnunen, A.K. Nuottaj\"arvi, J.P. Pekola, and W.M. Bergmann Tiest}
\label{fig:alphas}
\end{center} 
\end{figure}

We first carried out an $R-T_0$ measurement in a dilution refrigerator  measuring the resistance $R$ of the CorTES using a 4-wire AC 
method with a current bias of 5 $\mu$A as a function of $T_0$. From this, $T_{\rm c}=123$ mK was obtained. Next, we measured a set 
current-voltage $[I(V)]$ curves using voltage bias with a source impedance $R_{\rm s}$ of 7 m$\Omega$, and a SQUID ammeter. The 
results are shown in Fig. \ref{fig:cortes_fig1}, together with a fit using the model above. The $I(V)$ curve is insensitive to the 
external magnetic field thanks to the Nb groundplane. The fitting parameter is $\Lambda$ in the the superconducting region. Best fit 
yields $\Lambda=1.25k_{\rm B} T_{\rm c}$, which is a reasonable value, somewhat smaller than the BCS gap, $\Delta=1.76 k_{\rm 
B}T_{\rm c}$. The sharp corner present in the fit is due to the fact that the model assumes a step-wise transition with zero width, 
whereas the actual transition is smooth, as seen in the inset of Fig. \ref{fig:alphas}. Figure \ref{fig:alphas} shows the steepness 
of the transition, $\alpha=d \ln R/d \ln T$, as a function of $V$ measured at different bath temperatures. In all the curves 
$\alpha$ has a maximum value of about 300. This is one order of magnitude higher than in typical (square) microcalorimeters 
\cite{wouter1}. The high $\alpha$ is attributed mainly to the self-screening property of the ground plane and the well defined edge 
conditions.

The noise characteristics of the CorTES were measured as a function of the bias voltage. Both noise spectra and the rms current 
noise between 100 Hz and 20 kHz were determined. The intrinsic thermal time constant of the CorTES with heat capacity $C$ and  
thermal conductance to the bath of $G$, $\tau_0=C/G$, was determined from pulse response to be about 1.2 ms. Thus, the rms 
measurements are mainly sensitive to noise which is not suppressed by the electro-thermal feedback (frequencies above $(2 \pi 
\tau_0)^{-1}$=130 Hz). This method allows us to investigate the current noise against the operating point. When biased in the 
operating region ($V\lesssim$ 1 $\mu$V), the noise in the CorTES can not be accounted for by assuming contributions from TFN, JN and 
SN, as can be seen in Fig. \ref{fig:cortes_fig4}. We argue that the discrepancy can not be explained by including an internal TFN 
(ITFN)  \cite{hoevers1,gildemeister2} arising from the finite internal thermal impedance of the TES film. The ITFN is calculated as 
in Ref. \cite{hoevers1} and it is proportional to  $I \alpha$ exhibiting a peak at a bias corresponding to the maximum value of 
$\alpha$. However, the ITFN does not explain the noise at lower bias voltages where the noise is more than a decade larger than what 
we would be expect just by assuming contributions from the previously known terms.

According to the Ginzburg-Landau theory, the free energy difference between equilibrium superconducting and normal states of a 
volume $\Omega$ in the absence of any fields is $F=\Omega (\alpha |\psi|^2 + 1/2 \beta |\psi|^4)$, where $\alpha=1.36 
\hbar^2/(4m\xi_0l)(T/T_{\rm c}-1) \equiv \alpha_0 (T/T_{\rm c}-1)$ and $\beta=0.108/N(0) [\alpha_0/(k_{\rm B} T_{\rm c})]^2$. Here 
$m$ is the electron mass, and $N(0)=1.33 \times 10^{34}$ cm$^{-3}$eV$^{-1}$ is the density of states at Fermi level for Ti, $\xi_0 
\approx 20$ $\mu$m is the BCS coherence length and $l\approx$ 100 \AA \; is the mean free path determined from the normal state 
resistivity. The temperature gradient within the normal section can be solved by $-\nabla(\kappa_{\rm N} \nabla T)=j(r)^2 \rho_{\rm 
n}=V^2L^{-1}[r \ln(r_{\rm b}/r_0)]^{-2}$ using boundary conditions $T(r_{\rm b})=T_{\rm c}$ and $\nabla T(r_0)=0$. Fluctuations of 
$\psi$ with free energy variations $\delta F \lesssim k_{\rm B}T$ are possible and they correspond to large fluctuating volumes 
$\delta \Omega$ of the condensate near $T_{\rm c}$. The temperature gradient restricts the fluctuations to an annular volume at the 
outer perimeter of the normal state part, where the coherence length $\xi(T)=0.86\sqrt{\xi_0 l)}/\sqrt{T/T_{\rm c}-1}$ diverges. We 
estimate the radial extent of fluctuations, $\delta r$, assuming that order parameter fluctuates between zero and its equilibrium 
value at a distance $\delta r$ away from the equilibrium phase boundary within a volume $\delta \Omega=2 \pi r_{\rm b} t \delta r$:
\begin{equation}
k_{\rm B} T_{\rm c} \simeq \delta F \simeq -\frac{\langle \alpha^2 \rangle}{2 \beta}\delta \Omega \simeq \frac{\pi \alpha_0^2 t 
\gamma^2 r_{\rm b}}{\beta} \delta r^3.
\label{eq:dF}
\end{equation}
Here we have assumed that $(T/T_{\rm c}-1)$ can be approximated by $\delta r \gamma$, where $\gamma T_{\rm c}$ represents the 
effective radial temperature gradient at the phase boundary. 
The solution for the temperature profile yields $\gamma=-V^2/(T_{\rm c}^2 aL r_{\rm b} \ln(r_{\rm b}/r_0) )$, where the numerical 
factor $0\le a\le 1$ is used as the only fitting parameter which describes the reduction of the Lorentz number close to the boundary 
due to the presence of Cooper pairs. Solving for $\delta r$ and inserting the equations for $\alpha_0$ and $\beta$, the fluctuation 
in boundary radius is given by $\delta r=0.48[\pi N(0) k_{\rm B} T_{\rm c} r_{\rm b} t \gamma^2]^{-1/3}$.

\begin{figure} 
\begin{center}
\includegraphics[width=12cm,angle=0]{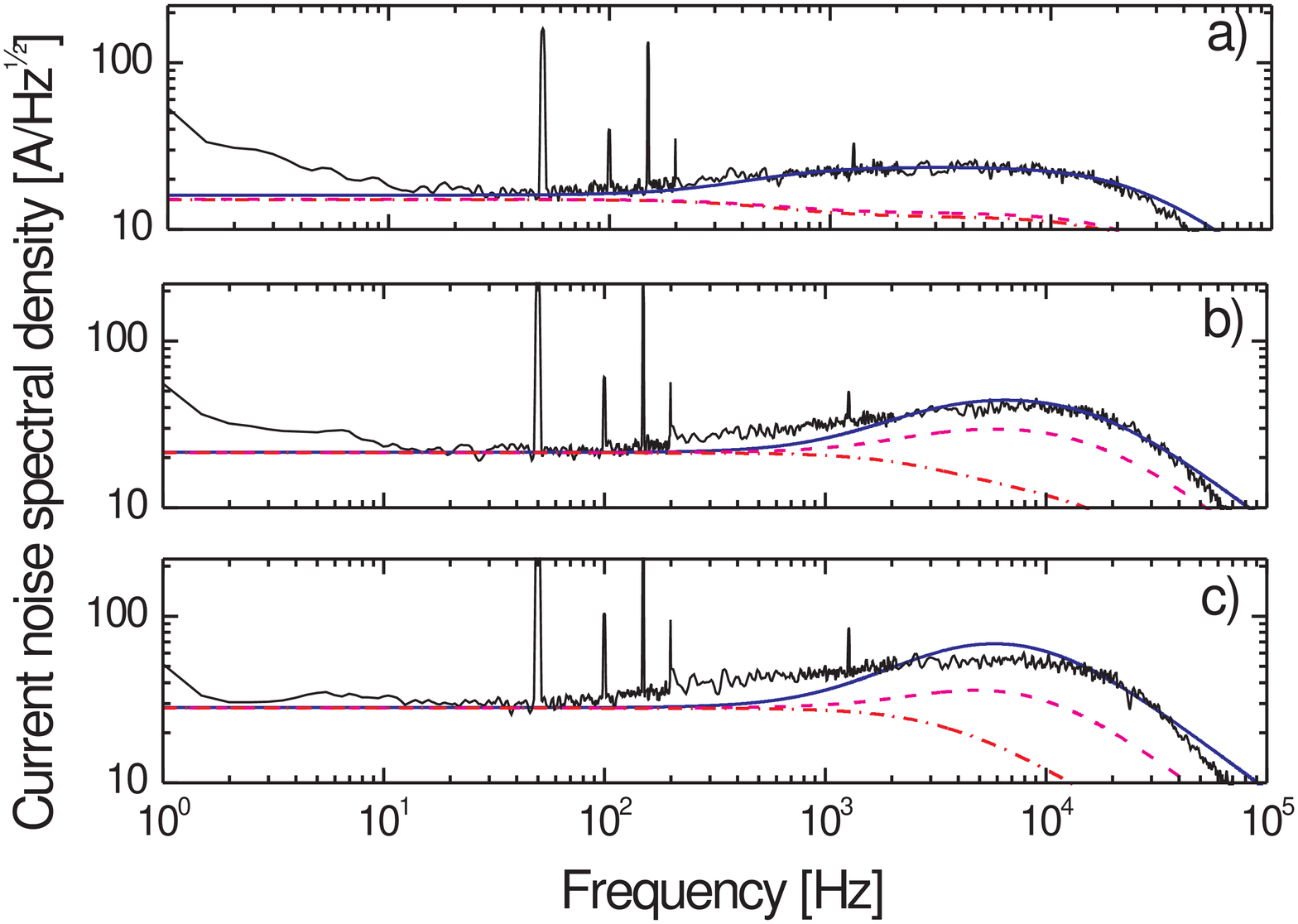}
\vspace{2cm}
\caption{A. Luukanen, H.F.C. Hoevers, K.M. Kinnunen, A.K. Nuottaj\"arvi, J.P. Pekola, and W.M. Bergmann Tiest}
\label{fig:spectra}
\end{center} 
\end{figure}

In order to obtain the spectral density of the critical fluctuations, we note that the relaxation time of a fluctuation is given by 
$\tau_{\rm GL}=\hbar \pi [8 k_{\rm B} (T-T_{\rm c}) ]^{-1}=\hbar \pi [8 k_{\rm B} T_{\rm c} \gamma \delta r ]^{-1}$. Thus, the 
equivalent noise bandwidth is $\int_0^{\infty} (1+\omega^2 \tau_{\rm GL}^2)^{-1} d\omega=\pi/(2 \tau_{\rm GL})$, and the resulting 
spectral density of the resistance fluctuations is given by $\delta R=\rho_{\rm n} \delta r (2 \pi t r_{\rm b})^{-1}\sqrt{2 
\tau_{\rm GL}/\pi}$. This can be considered as a white noise source within the bandwidth of our measurement. The current noise 
arising from resistance fluctuations is given by $\delta I=dI/dR \; \delta R = I/(2R) (b+1)\delta R$ where $R=V/I$ and $b=(R-R_{\rm 
s})/(R+R_{\rm s})$ corrects for non ideal voltage bias. The resistance fluctuations are suppressed by the ETF in a similar fashion 
as JN. More explicitly, the fluctuation superconductivity noise (FSN) component is given by
\begin{equation}
\delta I_{\mathrm FSN}(\omega)=\frac{I \delta R }{R}\frac{1+b}{2(1+ b L_0)} \frac{\sqrt{1+\omega^2 \tau_0^2}}{\sqrt{1+ \omega^2 
\tau_{\mathrm eff}^2}},
\label{eq:excs}
\end{equation}
where $bL_0=VI \alpha/(GT)$ is the loop gain of the negative electrothermal feedback, and $\tau_{\rm eff}=\tau_0/(1+ b L_0)$ is the 
effective time constant of the sensor.
 As Figs. \ref{fig:cortes_fig4} and \ref{fig:spectra} show, the noise can be accurately modelled through out the transition and over 
a wide range in frequency with only one fit parameter, $a=0.1$. We should, however, keep in mind that the crude definition of 
$\delta r$ in our model may simply be compensated by this fitting parameter.

When FSN dominates over the other noise terms, the FWHM energy resolution of a calorimeter can be estimated from the noise 
equivalent power NEP$_{\rm FSN}=\delta I_{\rm FSN}/S_I$ where $S_I$ is the current responsivity of the sensor \cite{moseley},
\begin{equation}
\Delta E \approx 1.18 \left( \int_0^{\infty} \frac{df}{{\rm NEP_{FSN}}^2(f)} \right)^{-1/2}=1.18 \frac{VI}{L_0 R} \delta R 
\sqrt{\frac{2 \tau_0}{\pi}} = 1.18 I^2 \frac{\delta R}{L_0}\sqrt{\frac{2 \tau_0}{\pi}}.
\label{eq:reso}
\end{equation}
In the measured device, $\Delta E_{\rm FSN}$ has a minimum value of $\sim$0.1 eV at $V=0.55$ $\mu$V (where $\alpha$ is at maximum) 
and it increases to 29 eV at $V=0.3$ $\mu$V. This implies that the energy resolution of TES microcalorimeters degrades significantly 
if biased at a voltage below that corresponding to maximum of $\alpha$.

In summary, we have fabricated and analyzed an idealized transition-edge sensor in which edge-effects are excluded. An analytical 
steady state model has been developed which shows good agreement with the measured $I(V)$ curve. The CorTES is insensitive to 
external magnetic fields due to a current carrying Nb ground plane. As a result, the $\alpha$ remains above 300 even when biased 
with constant voltage bias.  We show that the previously unexplained extra noise originates from thermal fluctuations of the phase 
boundary. The same noise mechanism is present in all types of superconducting transition-edge sensors \cite{nagaev}, but the effect 
might not be observable in some cases depending on the way the phase boundaries configure themselves. 

\acknowledgements

This work has been supported by the Academy of Finland under the Finnish Centre of Excellence Programme 2000-2005 (Project No. 
44875, Nuclear and Condensed Matter Programme at JYFL), and by the Finnish ANTARES Space Research Programme under the {\em High 
energy Astrophysics and Space astronomy} (HESA) consortium. The work of HFCH and WBT  is financially supported by the Dutch 
organisation for scientific research (NWO). The authors gratefully acknowledge K. Hansen, N. Kopnin and H. Sepp\"a for their 
comments.


\pagebreak
\section*{Figure legends}
\begin{figure} 
\begin{center}
\includegraphics[width=12cm,angle=0]{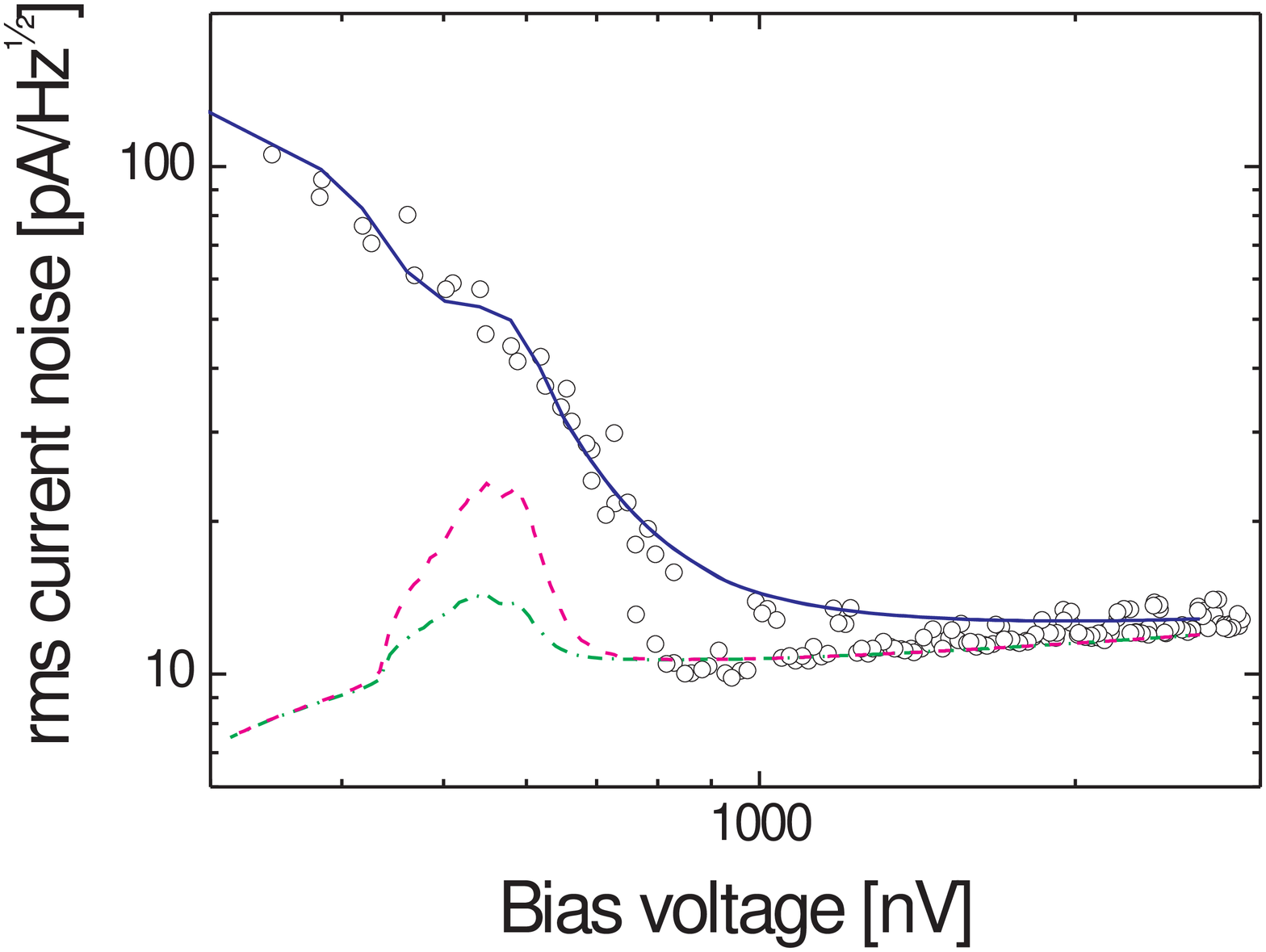}
\vspace{2cm}
\caption{A. Luukanen, H.F.C. Hoevers, K.M. Kinnunen, A.K. Nuottaj\"arvi, J.P. Pekola, and W.M. Bergmann Tiest}
\label{fig:cortes_fig4}
\end{center} 
\end{figure}

\begin{enumerate}
\item a) The measured $I(V)$ characteristics of the CorTES at different external magnetic fields denoted by the different symbols. 
The sensor itself is shielded by the Nb ground plane, and the $I(V)$ curve is not affected by the field. The Nb bias lines, however, 
are sensitive to the field and have a relatively small critical current due to difficulties in step coverage. The solid line 
corresponds to the fit with $\Lambda$=1.25$k_{\rm B}T_{\rm c}$. b) An optical micrograph of the sensor. The radii of the centre and 
outer contacts are 15 $\mu$m and 150 $\mu$m, respectively. c) A diagram showing the layer order.
\item Transition steepness, $\alpha$, measured at different bath temperatures. A lower bath temperature corresponds to a larger bias 
dissipation and correspondingly larger biasing currents. However, the relative transition width remains almost unchanged. The inset 
shows the corresponding $R-T$ curves calculated from the $I(V)$ curve.
\item The average rms current noise in a frequency band from 100 Hz to 20 kHz (marked by open circles) measured at a bath 
temperature of 20 mK. The total modeled noise is the solid line, and it consists of TFN, JN, SN, ITFN, and FSN. The dash-dotted line 
represents a model with only TFN, JN and SN included. The dashed line represents a model with TFN, JN, SN, and ITFN.
\item Noise spectra measured at a bath temperature of 20 mK at bias voltages of a) 0.75 $\mu$V, b) 0.62 $\mu$V, and c) 0.5 $\mu$V. 
The notation for the different modelled noise components is identical to that of Fig. \ref{fig:cortes_fig4}.
\end{enumerate}

\pagebreak

\pagebreak

\pagebreak

\pagebreak

\end{document}